\documentclass[letterpaper]{article}
\usepackage{cite}

\usepackage{helvet}
\usepackage{courier}

\usepackage{amssymb,amsfonts,amsmath}

\usepackage{bm}
\usepackage{multirow,rotating} 
\usepackage{color}
\usepackage{epsfig}

\usepackage{sgame}

\usepackage{authblk}

\newcommand{\ind}[1]{\ensuremath{\textbf{1}_{\{#1\}}}}

\renewcommand{\l}{\lambda}

\newtheorem{finding}{Finding}

\begin{document}
%

\title{Crowdsourcing Dilemma}


\author[1]{Victor Naroditskiy}
\author[1]{Nicholas R. Jennings}
\author[2]{Pascal Van Hentenryck}
\author[2]{Manuel Cebrian}
\affil[1]{University of Southampton}
\affil[2]{National Information and Communications Technology Australia}


\date{}

\maketitle

\begin{abstract}
Crowdsourcing offers unprecedented potential for solving tasks efficiently by tapping into the skills of large groups of people. A salient feature of crowdsourcing---its openness of entry---makes it vulnerable to malicious behavior. Such behavior took place in a number of recent popular crowdsourcing competitions. We provide game-theoretic analysis of a fundamental tradeoff between the potential for increased productivity and the possibility of being set back by malicious behavior. Our results show that in crowdsourcing competitions malicious behavior is the norm, not the anomaly---a result contrary to the conventional wisdom in the area. Counterintuitively, making the attacks more costly does not deter them but leads to a less desirable outcome. These findings have cautionary implications for the design of crowdsourcing competitions.
\end{abstract}

\section{Introduction}
\noindent
Numerous successful examples of the power of crowdsourcing to solve problems of extreme difficulty~\cite{howe2006rise,hand2010citizen,von2008recaptcha,von2006games,horowitz2010anatomy,huberman2009crowdsourcing,cooper2010predicting,pickard2011time,mason2012conducting,rahwan2012global,barrington2012game,hellerstein2011searching,alstott2013predictors,zhang2012task} have overshadowed important episodes where elaborate sabotage derailed or severely hindered collective efforts. The winning team in the DARPA Network Challenge obtained the locations of the ten balloons after spending significant efforts filtering the majority of false submissions, including fabricated pictures containing individuals in disguise impersonating DARPA officials~\cite{tang2011reflecting}. A team from the University of California at San Diego lost its lead in the DARPA Shredder Challenge after their  progress was completely wiped-out by a relentless number of coordinated overnight attacks~\cite{cebriannas,shredder}. The team that topped the US Department of State sponsored Tag Challenge had to withstand a smear campaign orchestrated in Twitter aimed at reducing its credibility~\cite{rahwan2012global}. Beyond crowdsourcing competitions, Ushahidi's collective conflict mapping for the Arab Spring had to be shut down for long periods of time  due to suspicions that it had been infiltrated by government officials~\cite{cebriannas}. This fragility of crowdsourcing can also be used in a positive manner: targeted attacks can exploit the fragility of decentralized, hierarchical recruitment of on-line extremists, effective slowing down their radicalization and training process~\cite{extremistfora2013}.


These episodes have received a response in the form of emerging work at the intersection of the computer and economic sciences. Recent results in this area have elucidated that it is possible to design incentive structures and algorithmic strategies to modify the efficacy and quality of the crowdsourced solution, the vulnerability of crowdsourcing to malicious behavior, and the cost of undertaking it~\cite{kleinberg2005query,kittur2013future,zhang2011crowdsourcing,mao2011human,karger2011budget,bernstein2012analytic,ipeirotis2011crowdsourcing,kamar2012incentives,ipeirotis2011managing,lorenz2011social,pfeiffer2012adaptive,mason2012collaborative,naroditskiy2012verification,cebrian2012finding,tran2012efficient,venanzi2013trust,babaioff2012bitcoin,rutherford2013limits,woolley2010evidence,pentland2012new,drucker2012simpler,andersonsteering,nath2012threats,chitnis2012game}.  
These three factors can attain a wide variety of  values depending on the particularities of the problem at hand, the economic incentives at stake, as well as the algorithmic platform supporting the collective effort. For instance, the combinatorial nature of puzzle-assembly in the DARPA Shredder Challenge makes the solution very vulnerable to an attack at a low cost---destroying puzzle progress is much easier than creating it. On the other hand, fabricating a false balloon sighting in the DARPA Balloon Challenge is costly, as it involves a certain degree of counterintelligence skills, more so when individuals posing in false submissions risk identifiability---while at the same time, a false submission does not affect the validity of the other submissions. 

In this paper, we propose a formal analysis to explore the efficacy-vulnerability tradeoff of crowdsourcing. We adopt a  scenario where two firms (players) compete against each other to obtain a better solution to a task which can be crowdsourced. Our goal is to understand whether malicious behavior exemplified above is the norm or the anomaly, i.e. whether we should expect players to undertake attacks on the collective progress of competing players. We consider  the cost of attacking a crowdsourcing strategy, and investigate the effect of higher costs on behavior of the players. We also aim at quantifying how this malicious behavior affects the likelihood of using crowdsourcing as a strategy, and ultimately how it impacts the social welfare under collective problem solving. Such an understanding will be helpful for the design of future crowdsourcing competitions, as well as support the decision process of institutions and firms considering crowdsourcing.

Our main finding is that making attacks more costly perhaps by making it more difficult to attack, does not deter the attackers and results in a more costly and less efficient equilibrium outcome. Paradoxically, making the cost of attacking zero is best for the players. This suggests that care should be used when attempting to discourage attacks by raising the cost. In situations where damage from attacking is high, the incentive to attack is very strong. Instead, crowdsourcing competitions are more suited for scenarios where damage inflicted by an attack is low.

In the next section, we present a model for studying the tradeoff between higher productivity offered by crowdsourcing and the increased vulnerability that comes with it. Then, we derive  equilibrium of the game defined by the model. We pay particular attention to the case where both players choose to crowdsource and investigate the players' incentives to attack. We also analyze how the cost of an attack and the damage from the attack affect the incentives of players to crowdsource.



\section{Model}
\noindent
We study a non-cooperative situation where two players (or firms) compete to obtain a better solution to a given task. The firm with the better solution wins and receives a reward of $R$. Each firm can develop an in-house solution or crowdsource the task. The former is referred to as the closed strategy,  the latter---as open. We focus on scenarios where open strategies are likely to be more efficient even though the exact level of efficiency is not known until after a firm engages in crowdsourcing---this reflects a high level of uncertainty underlying engagement processes in social networks~\cite{centola2007complex,liben2008tracing,golub2010using,bakshy2011everyone,iribarren2009impact,goyal2012competitive,toole2012modeling,ugander2012structural}. 
Also, open strategies are susceptible to attacks: a firm using an open strategy can be attacked by the competing firm. The resulting damage from the attack impairs the other firm, and may let the attacking firm win.

To investigate the tradeoff between higher efficiency of crowdsourcing and  vulnerability to attacks, we propose a model that isolates these two factors. In our model, a firm decides whether or not to crowdsource, and whether or not to attack the competitor. The nature of a crowdsourcing strategy is that it is  observable by everyone including the competitors. If the competitor chooses to crowdsource, the opponent would observe it, and can decide whether or not to attack. Notice that the decisions are sequential: the decision about crowdsourcing is made first, while the decision about attacking is made second. We will model this interaction as a sequential game.

Formally, we have a two-stage game. In the first stage, the firms decide whether to solve the task in-house $S$ or to crowdsource it $C$. Then the efficiencies of the firms that  chose to crowdsource are observed. Due to the open nature of crowdsourcing, the efficiency, or {\em productivity}, of a crowdsourcing strategy used by player $i$ becomes known not just to the crowdsourcing player but also to the opponent. We use $P_i$ for the random variable denoting the productivity and $p_i$ for its realized value. The productivity of the in-house solution is fixed and normalized to zero. We assume that the productivity of a crowdsourcing strategy is uniformly distributed between 0 and 1: $P_i\sim U[0,1]$. In the absence of attacks, the firm with a higher  productivity wins.

The assumption that the productivity of a crowdsourcing firm is publicly known is  consistent with competitions such as the Network Challenge and the Shredder Challenge. In the Shredder Challenge, the current progress of each competitor was publicly known. In the Network Challenge, the effectiveness of a crowdsourcing strategy could be estimated through the social media impact generated by a crowdsourcing team.

In the second stage, the players decide whether or not to attack the opponent (attacking is denoted by $A$ and not attacking by $N$). An attack is costly, and the cost $q \in (0,1)$ is expressed as a fraction of the total reward $R$. This cost can represent a range of situations: the human effort in disrupting the opponent's solution; the complexity of creating multiple identities to carry out a Sybil attack;  financial punishment received when the attack is detected in a crowdsourcing competition. The damage inflicted by the attack is denoted by $d \in (0,1)$, which determines how much productivity is taken away from the open strategy (equivalently, how much more productive the attacking firm becomes after ``stealing" the crowdsourced solution). The firm that has a higher productivity at the end of the second stage wins the prize $R$, which is normalized to be $R=1$. The parameters $q$ and $d$ are publicly known. We characterize the equilibrium of this two-stage game as a function of these parameters.

\section{Equilibrium Analysis}
\noindent
We find the subgame perfect equilibrium of the two-stage game. Each pair of players' decisions made in the first stage (to crowdsource or not) result in a different second-stage game. We first analyze each second-stage game (to attack or not). Based on these, we then analyze how decisions are made in the first stage.

When both players use an in-house strategy $S$, there is no reason to attack, and they both choose $N$ in the second stage. Each one is equally likely to win and the expected utility of each is $\frac{1}{2}$ . We note this in the $SS$ cell of the payoff matrix in Table~\ref{fig:fullgame}. 

When player 1 crowdsources and player 2 does not, player 1 has no reason to attack, but player 2 attacks if doing so puts him ahead of player 1. Player 2 attacks if the realized productivity of player 1 is less than the damage $p_1 < d$ player 2 can inflict. Recall that we normalized productivity of an in-house solution to zero. The ex-ante utility of player 1 before his productivity is realized is $1-d$: i.e., he receives the payoff of 1 when his productivity is high enough to not be attacked, which, for uniformly distributed productivity, happens with probability $1-d$. 
The ex-ante utility of player 2 before the productivity of player 1 is realized is $d(1-q)$: i.e., the productivity of player 1 is low enough to be overtaken after an attack (which happens with probability $d$), so player 2 attacks and receives the reward minus the cost: $1-q$.  Note that player 2 attacks for any cost of an attack $q \in [0,1)$: attacking brings a positive utility while not attacking results in zero utility. 
The case when player 2 crowdsources, and player 1 does not is symmetric. We summarize this in the $CS$ and $SC$ cells of the payoff matrix in Table~\ref{fig:fullgame}. 

The most interesting case is when both players crowdsource. Let $p_1$ and $p_2$ denote their productivities, which are known before they decide on attacking. Consider the case when $p_1 > p_2$. The other case is symmetric. 

If the difference in productivities $p_1-d > p_2$ of the players is so large that the attack by player 2 does not let her reach player 1, then attacking does not change the outcome and neither player attacks. In this case, player 1 receives the utility of 1, while player 2 receives the utility of 0.  

\subsection{Crowdsourcing and Attacking}
\noindent
We analyze the game when both players crowdsource and a unilateral attack by the weak player (i.e., player 2) will bring him ahead of the strong player (i.e., player 1). This is the case when $p_1-d<p_2$. In this case, player 2 would like to attack if player 1 does not. At the same time, player 1 would like to attack in order to keep its lead only if player 2 is attacking. Consequently, there is no pure strategy equilibrium of this game. The payoff matrix showing utilities for each player appears below.

\begin{center}
\begin{game}{2}{2}
  {}    & $A$    & $N$\\
$A$   &$1-q,-q$   &$1-q,0$\\
$N$   &$0,1-q$   &$1,0$
 \end{game}
\end{center}

The game possesses a unique mixed equilibrium. Let $\l_1,\l_2$ denote the probabilities that, respectively, player 1 and 2 attacks. For player 1 attacking results in the expected utility of 
\begin{align}
& \l_2(1-q) + (1-\l_2)(1-q) = 1-q\label{eq:u1a}
\end{align}
while non-attacking gives the expected utility of 
\begin{align}
\l_2 \cdot 0 + 1 \cdot (1-\l_2) = 1-\l_2 \label{eq:u1b}.
\end{align}
In a mixed equilibrium, a player's expected utility from choosing either action must be the same. This is satisfied for player 1 when 
\begin{align}
& 1-q = 1 -\l_2 \label{eq:c1}
\end{align}
that is for $\l_2=q$. Similarly, for player 2 the expected utility from attacking is 
\begin{align}
& \l_1(-q) + (1-\l_1)(1-q) = 1-q-\l_1\label{eq:u2a}.
\end{align}
The expected utility from not attacking is $0$, yielding the equilibrium condition 
\begin{align*}
& \l_1=1-q .
\end{align*}
Thus, in the mixed equilibrium, player 1 attacks with probability $\l_1=1-q$, and player 2 attacks with probability $\l_2=q$ leading to the following observation.
\begin{finding}
The higher the cost, the more likely the weak player (i.e., player 2) is to attack and the less likely is the strong one. 
\end{finding}

This behavior contradicts the intuition that making attacks more costly helps prevent them: that is, the higher it is to attack, the less should either player attack. We explain why this does not hold by looking into the reasons for players' attacks. Player 1 attacks only to counteract a possible attack of player 2. Player 2 attacks in the hope that the attack is not counteracted by player 1 allowing player 2 to get ahead. Crucially, the incentive of player 1 to attack {\em increases} in the likelihood that player 2 attacks, while the incentive of player 2 to attack {\em decreases} in the likelihood that player 1 attacks. 

The cost of attacks {\em increases} the likelihood that player 2 attacks because player 1 attacks less when the cost is higher. Whenever both players attack, player 1 wins with certainty, and player 2 would have preferred to avoid the useless and costly attack. Thus, the less player 1 is likely to attack (i.e., the higher the cost), the more eager player 2 is to attack. For example, when attacking is free, player 1 always attacks and always wins obtaining the utility of 1. When the cost is the same as the prize, $q=1$, player 2 always attacks taking the prize away from player 1 who does not attack resulting in zero utility for both.  This means that the higher the cost of an attack, the more likely is the weak player to win! 
\begin{finding}
\label{fn:weaker}
Instead of protecting the better crowdsourcing strategy, a higher cost of an attack results in the weaker strategy winning after a unilateral attack.
\end{finding}

This is a striking observation. Making attacks more costly should help the society by ensuring the stronger strategy wins. This does not occur as, due to the higher likelihood of agent 2 attacking, the weaker player attacks and wins. Furthermore,  the players spend more resources on attacking each other: making attacks more costly results in a lower  total utility of the agents. To show this, we first notice that the expected number of attacks does not change with the cost of an attack. Raising the cost result in fewer attacks by the strong player and in more attacks by the weak player: the weak player attacks with probability $q$ and the strong player attacks with probability $1-q$. 
\begin{finding}
\label{fn:samecost}
The expected number of attacks is one regardless of the cost of an attack $q$.
\end{finding}
A direct consequence is the following:
\begin{finding}
\label{fn:welfare}
Increasing the cost of an attack decreases (not increases) the total utility of the agents.
\end{finding}
We now look at the competition from the point of view of what is socially optimal. We measure the benefit to the society as the total utility of the players and the final quality of the solution. We refer to it as {\em social welfare}. Combining Findings \ref{fn:weaker} and \ref{fn:welfare}, we conclude that:
\begin{finding}
\label{fn:lowerwelfare}
Increasing the cost of an attack results in a lower social welfare.
\end{finding}
This finding suggests that increasing the cost of an attack is damaging rather than helping. Setting the costs to zero would be optimal from the social welfare point of view: the strong player would attack and win with probability 1, the weak player would not attack, and the cost of attacking would be zero.
\begin{finding}
\label{fn:zero}
Free attacks maximize the social welfare.
\end{finding}



We now turn to the analysis of players' utilities. From \eqref{eq:u1a} or \eqref{eq:u1b}, the expected utility of the strong player is $1-q$, while from~\eqref{eq:u2a}, the expected utility of the weak player is $0$. Observe the equilibrium strategy requires that the weak player obtains the same utility from both actions: i.e., the strong player selects her action so that attacking by the weak player gives exactly zero expected utility. 


\subsection{To Crowdsource or Not}
\noindent
The expected utilities discussed above are for the game played in the second stage after productivities become known and for the case when $p_1-d <p_2$. 


We now take a step back and compute expected utilities when both players crowdsource but before the productivities become known. To avoid confusion we refer to these utilities as {\em ex-ante} utilities. The ex-ante utility of player 1 (and symmetrically of player 2) is 
\begin{align*}
& u_1 = \Pr(P_2 < P_1 < P_2+d)(1-q) + \Pr(P_2+d < P_1) .
\end{align*}
The first term corresponds to the utility of player 1 in the mixed equilibrium of the game described above, and the second term corresponds to player 1 winning with certainty when player 2 cannot reach him even after attacking. Under the assumption that $P_1$ and $P_2$ are uniformly distributed between zero and one, we derive
\begin{align*}
& \Pr(P_2 < P_1 < P_2+d) = \int_{0}^1\int_{0}^1 \ind{p_2 < p_1 < p_2+d} dp_2dp_1 \\
& = \int_0^1 \int_{p_2}^{\min(p_2+d,1)}1dp_1dp_2\\
& = \int_{0}^1 (\min(p_2+d,1) - p_2) dp_2 \\
& = \int_0^{1-d} d\ dp_2 + \int_{1-d}^1 (1-p_2)dp_2 \\
& = (1-d)d + (1-\frac{1}{2}) - (1-d - \frac{(1-d)^2}{2}) = d - \frac{d^2}{2}\\
& \Pr(P_2+d < P_1) = \int_0^1\int_0^1 \ind{p_2+d<p_1}1 dp_2dp_1\\
& = \int_d^1 \int_0^{p_1-d}1 dp_2dp_1 = \int_d^1 (p_1-d)dp_1 \\
& = \frac{1}{2} - d - (\frac{d^2}{2} - d^2) = \frac{d^2}{2} - d + \frac{1}{2} 
\end{align*}
The ex-ante utility of either player is
\begin{align*}
& u_1 = u_2 = (d-\frac{d^2}{2})(1-q) + (\frac{d^2}{2}-d+\frac{1}{2})1 \\
& = \frac{1}{2} - (d-\frac{d^2}{2})q
\end{align*}

The utility of the players decreases in both $q$ and $d$. At the extreme case when both $q$ and $d$ are 1, the utility is zero. Whenever either of the parameters is at its minimum value of zero, the utility is at its maximum value of $\frac{1}{2}$. Observe that the ex-ante utility decreases in the cost of an attack. This is a consequence of the number of attacks being independent of $q$ when both $p_1-d<p_2$ as we noted in Finding \ref{fn:samecost} and a related Finding \ref{fn:welfare}.

In the Prisoner's Dilemma~\cite{rapoport1965prisioner,trivers1971evolution,aumann1981survey,axelrod1981evolution}, the unique equilibrium for both players is to choose the action that hurts the other player: i.e., to defect. In the resulting equilibrium, both players are hurt. A similar situation (although in mixed strategies) arises in the crowdsourcing game. Both players choose to attack (in equilibrium, the expected number of attacks is 1), incurring unproductive costs. When both players attack, the outcome is the same as when neither player attacks except that each player incurs the cost of an attack. When only the strong player attacks, the outcome does not change, but the player incurs the cost of an attack. When only the weak player attacks, the outcome changes for a less efficient outcome (the weaker player wins), and the weak player incurs the cost of an attack. Only when neither player attacks, there is no loss to social welfare.

Having analyzed the second stage games, we can now describe the entire payoff matrix for the game played in the first stage. Note that in the first stage, the players are unaware of their productivities, should they choose crowdsourcing. Thus, the payoff matrix contains their ex-ante utilities, depicted in Table~\ref{fig:fullgame}.

\begin{table*}
\begin{center}
\begin{game}{2}{2}
      & $C$    & $S$\\
$C$   &$\frac{1}{2} - (d-\frac{d^2}{2})q, \frac{1}{2} - (d-\frac{d^2}{2})q$   &$1-d,d(1-q)$\\
$S$   &$d(1-q),1-d$   &$\frac{1}{2},\frac{1}{2}$
\end{game}
\end{center}
\caption{{\bf Expected payoff matrix for the crowdsourcing game.}}
\label{fig:fullgame}
\end{table*}

We ask the question of how $q$ and $d$ affect the equilibria. For any choice of parameters, only $CC$ and $SS$ can be pure strategy equilibria. Higher damage from an attack and low cost of attacking correspond to the crowdsourcing strategy being more risky. Indeed, as we detail below, for $d>\frac{1}{2}$ and $q<\frac{2d-1}{d^2}$, $SS$ is the only equilibrium strategy. Similarly, when the damage is low and cost of attacking is high, crowdsourcing dominates, and $CC$ is the only equilibrium strategy for  $d<\frac{1}{2}$. For the remaining range of parameters $d \ge \frac{1}{2}$ and $q\ge\frac{2d-1}{d^2}$, both  $CC$ and $SS$ are equilibrium strategies.

Crowdsourcing $CC$ is a unique equilibrium when the damage inflicted by an attack is low $d< \frac{1}{2}$. Indeed, when both firms use closed strategies, deviating to a crowdsourcing strategy provides a higher payoff  ($1-d > \frac{1}{2}$), as in expectation the productivity of a crowdsourcing strategy will make the crowdsourcing firm outside the reach of the in-house competitor even after being attacked. Crowdsourcing for one firm and not crowdsourcing for the other is not stable, as the non-crowdsourcing firm is better off switching to crowdsourcing $\frac{1}{2} - (d-\frac{d^2}{2})q > d(1-q)$. This holds regardless of the cost of an attack $0< q <1$. 

For $d\ge \frac{1}{2}$, $CC$ remains an equilibrium only if the cost of an attack is high enough: $q\ge\frac{2d-1}{d^2}$. 
Intuitively, players crowdsource when attacking costs a lot (higher $q$) and is not effective (lower $d$). 


In summary, crowdsourcing $CC$ is the unique equilibrium when $d<\frac{1}{2}$; in-house $SS$ is the unique equilibrium for $d > \frac{1}{2}$ and  $q<\frac{2d-1}{d^2}$; both $CC$ and $SS$ are equilibrium strategies for $d \ge \frac{1}{2}$ and $q\ge\frac{2d-1}{d^2}$.

How does the cost of an attack influence the likelihood of players to crowdsource? Intuitively, we would expect that high costs prevent attacks and make crowdsourcing more appealing. Our model provides reasoning against this intuition. The analysis above reveals that $q$ has a limited affect on the choice of the equilibrium strategy and for the case when both players crowdsource, it does not reduce the expected number of attacks. Cost of attacking matters when the level of damage is high ($d\ge \frac{1}{2})$. In this case, high costs enable $CC$ as equilibrium, however $SS$ still remains a equilibrium.

The damage from an attack has a stronger effect on the equilibrium. Low damage corresponds to fewer attacks on a crowdsourcing firm: when both players crowdsource the likelihood that the strong player is within reach of the weak player is proportional to the level of damage, $\Pr(P_2 < P_1 < P_2+d)$. Crowdsourcing is the unique equilibrium strategy for low levels of damage ($d<\frac{1}{2})$. This leads to a  conclusion that competitions where the potential to inflict damage on a crowdsourcing opponent is low are more likely to promote crowdsourcing than competitions where a high costs of attacking are used to deter attack.


\section{Discussion}
\noindent
Our results bear resemblance to the Prisoner's Dilemma but paint an even starker picture.  When both players crowdsource (i.e., choose a more efficient way of performing the task) and are close to each other (specifically, within the damage inflicted by attacking, as was the case in the DARPA Shredder Challenge) in terms of solution quality, the expected number of attacks is one regardless of the cost of an attack. Increasing the cost of an attack offers no deterrence. Therefore, under our basic model, malicious behavior is the expected behavior, not the anomaly. Given this result and the examples of malicious behavior in competitions,  more emphasis should be given to the issue. There has been significant academic interest towards filtering misinformation, however models of malicious behavior  in crowdsourcing scenarios have been absent until this work.

The finding that raising costs of attack is harmful for the players is striking and warrants further empirical investigation. Our model predicts that higher costs of attacks lead to more attacks by the weak player resulting in a higher probability that the weak player would win. Furthermore, the expected number of attacks remains the same, resulting in a higher costs incurred by the players on orchestrating attacks. Confirming these findings in the lab or fiend experiment is a direction for future work.

We made a number of modeling choices: two firms, perfect observability of productivities, risk-neutrality of the agents, and provided the analysis for a uniform distribution of productivities. These choices were guided by the simplicity and the goal to isolate the factors relevant to the tradeoff between higher productivity of open strategies and higher vulnerability. Due to the simplicity of our model, we believe that our results capture the fundamental features of the tradeoff between the productivity/vulnerability.

We considered the relative performance of the firms while ignoring the absolute quality of the solution. This explains why the highest expected social welfare is obtained when neither firm crowdsources, and therefore, neither firm attacks (see the payoff of the $(S,S)$ strategy in Table~\ref{fig:fullgame}).  This is suitable for many competition settings where only relative performance is important (such as the aforementioned DARPA Network Challenge, DARPA Shredder Challenge, or Tag Challenge). Requiring a certain minimum solution quality and modeling the cost of effort are interesting extensions for future work. For example, one could model effort as time required to find a solution, with the time being inversely related to productivity (e.g., $t = 1-p$). Competitions may also mitigate aggression by using reward mechanisms where the reward received by the winner depends on the global progress of all teams---linking crowdsourcing games to Public Good Games~\cite{fehr1999theory,rand2009positive,fowler2005altruistic,fowler2010cooperative}.  

Repeated encounters in crowdsourcing competitions may provide opportunities for the emergence of a richer set of socially desirable strategies as in the Iterated PrisonerÕs Dilemma~\cite{fudenberg1986folk,fundenberg1990evolution,nowak1993strategy}. It would also be interesting to study how the presence of more than two players affects the behavior displayed. 

Our results emphasize that despite crowdsourcing being a more efficient way of accomplishing many tasks, it  also a less secure approach. In scenarios of  ``competitive'' crowdsourcing, where there is an inherent desire to hurt the opponent, attacks on crowdsourcing strategies are essentially unavoidable. We expect these surprising results derived in our stylized model to hold in a variety of more complicated scenarios that exhibit the fundamental tension between openness, efficiency and vulnerability.\\





\end{document}